\documentclass[final]{svjour2}
\usepackage{graphicx}
\usepackage{rotating}
\usepackage{amssymb}
\usepackage{mathptmx}
\usepackage[numbers]{natbib}
\makeatletter
\journalname{Journal of Low Temperature Physics}

\bibpunct{}{}{,}{s}{}{,}

\begin{document}

\newcommand{\hdblarrow}{H\makebox[0.9ex][l]{$\downdownarrows$}-}
\title{Latest NIKA results and the NIKA-2 project}

\author{A. Monfardini$^1$ \and R. Adam$^2$ \and A. Adane$^3$ \and P. Ade$^4$ \and P. Andr\'e$^5$ \and A. Beelen$^6$ \and B. Belier$^7$ \and A. Benoit$^1$ \and A. Bideaud$^4$ \and N. Billot$^9$ \and O. Bourrion$^2$ \and M. Calvo$^1$ \and A. Catalano$^2$ \and G. Coiffard$^3$ \and B. Comis$^2$ \and A. D'Addabbo$^1$ \and F.-X. D\'esert$^8$ \and S. Doyle$^4$ \and J. Goupy$^1$ \and C. Kramer $^9$ \and S. Leclercq $^3$ \and J. Macias-Perez$^2$ \and J. Martino$^6$ \and P. Mauskopf$^4$ \and F. Mayet$^2$ \and F. Pajot$^6$ \and E. Pascale$^4$ \and N. Ponthieu$^8$ \and V. Rev\'eret$^5$ \and L. Rodriguez$^5$ \and G. Savini$^{10}$ \and K. Schuster$^3$ \and A. Sievers$^9$ \and C. Tucker$^4$ \and R. Zylka$^9$}

\institute{1 Institut N\'eel, CNRS and Universit\'e de Grenoble, France\\ 
2 Laboratoire de Physique Subatomique et de Cosmologie (LPSC), CNRS and Universit\'e de Grenoble, France\\ 
3 Institut de RadioAstronomie Millimetrique (IRAM), Grenoble, France\\ 
4 Astronomy Instrumentation Group, University of Cardiff, UK\\ 
5 CEA-Irfu, Saclay, France\\ 
6 Institut d'Astrophysique Spatiale (IAS), CNRS and Universit\'e Paris Sud, Orsay, France\\
7 Institut d'Electronique Fondamentale (IEF), Universit\'e Paris Sud, Orsay, France\\
8 Institut de Plan\'etologie et d’Astrophysique de Grenoble (IPAG), CNRS and Universit\'e de Grenoble, France\\
9 Instituto Radioastronom\'ia Milim\'etrica (IRAM), Granada, Spain\\ 
10 University College London (UCL), UK\\
\email{monfardini@grenoble.cnrs.fr}
}

\date{15.07.2013}

\maketitle

\begin{abstract}

NIKA (New IRAM KID Arrays) is a dual-band imaging instrument installed at the IRAM (Institut de RadioAstronomie Millimetrique) 30-meter telescope at Pico Veleta (Spain). Two distinct Kinetic Inductance Detectors (KID) focal planes allow the camera to simultaneaously image a field-of-view of about 2 arc-min in the bands 125 to 175 GHz (150 GHz) and 200 to 280 GHz (240 GHz).
The sensitivity and stability achieved during the last commissioning Run in June 2013 allows opening the instrument to general observers.
We report here the latest results, in particular in terms of sensitivity, now comparable to the state-of-the-art Transition Edge Sensors (TES) bolometers, relative and absolute photometry. 
We describe briefly the next generation NIKA-2 instrument, selected by IRAM to occupy, from 2015, the continuum imager/polarimeter slot at the 30-m telescope. 


\end{abstract}

\section{Introduction}

Millimetre astronomy, traditionally important for the study of the early/cold evolutionary stages of the main astrophysical objects, i.e. stars, galaxies, is now a central subject for cosmological studies too. This interest has been further boosted by the recent Planck multi-band (22 GHz to 850 GHz) surveys\cite{planck}. In addition, the Herschel space telescope, operated at sub-millimetre and far-infrared wavelengths, mapped efficiently large portions\cite{herschel} of the Sky discovering a huge number of potentially interesting follow-up sources for a large ground-based telescope operating at longer wavelenghts. 
The 30-m IRAM telescope (Pico Veleta, Spain, 2850 meters a.s.l.) is, with its 6.5 arc-min field-of-view (FoV) and angular resolution of 10 arc-sec at 240 GHz (16 arc-sec at 150 GHz), a crucial tool to bridge, in terms of sensitivity and angular scale, between the ground-based interferometers (e.g. ALMA\cite{alma} and NOEMA\cite{noema}, high angular resolution, high sensitivity, small field-of-view) and the small Space telescopes (e.g. Planck, coarse angular resolution, moderate limiting sensitivity, surveys). The IRAM telescope is already equipped with state-of-the-art heterodyne instruments performing high-resolution spectroscopy between 74 GHz and 360 GHz. A large field-of-view, multicolor continuum instrument able to detect fainter sources is the natural complement to the existing instrumentation. 

\begin{figure}
\begin{center}
\includegraphics[%
  width=0.7\linewidth,
  keepaspectratio]{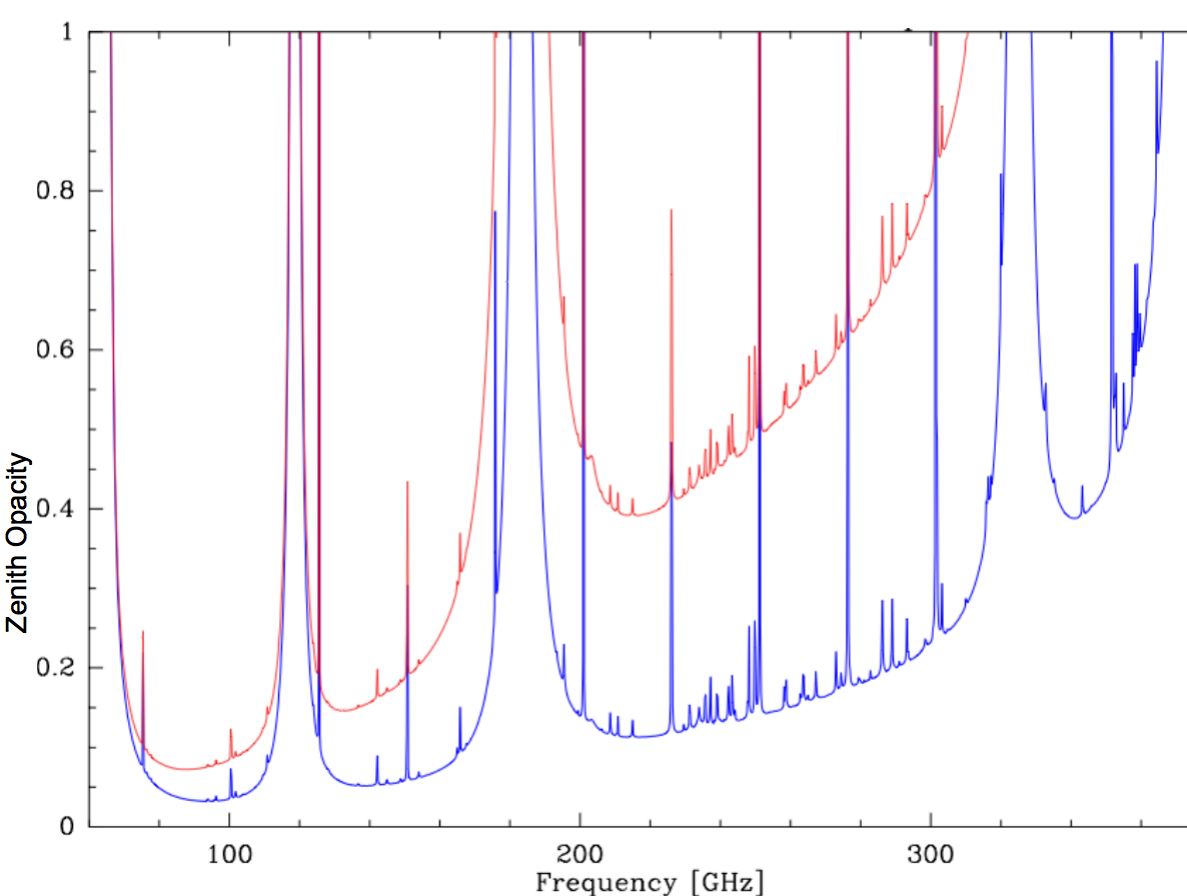}
\end{center}
\caption{(Color online) Atmosphere opacity (zenith) at Pico Veleta. Blue: 2 mm precipitable water vapor (pwv), good observing conditions; Red: 7 mm pwv, average summer conditions.}
\label{IRAM_sky}
\end{figure}

As shown in Fig.~\ref{IRAM_sky}, showing the atmpsphere opavity as a function of frequency, the Pico Veleta is an excellent site for observations at millimetre wavelengths, well suited in particular for the atmospheric windows near 150 GHz and 240 GHz.
The two continuum cameras on site, GISMO\cite{gismo13} and NIKA are operating at these very frequencies (GISMO at 150 GHz, NIKA at 150 and 240 GHz).

To preserve the good angular resolution and entirely fill the available FoV, a large number (thousands) of detectors (pixels) are needed. For this reason, and taking into account the limited manpower/financial resources and the need for a rapid development, the KID technology seems naturally adapted. 

Besides NIKA, KID are adopted in the planned submillimetre instruments MAKO\cite{swenson13}, MUSIC\cite{music13} and A-MKID\cite{amkid13}. In the visible/NIR, KID are successfully operating in the ARCONS\cite{Mazin13} camera installed at the Palomar 200-inches telescope.

\section{The NIKA development: 2009-2013}
The NIKA program started in November 2008, mainly driven by the new technology developments ongoing on the Kinetic Inductance Detectors (KID) at SRON, Cardiff and Institut N\'eel, Grenoble. The goal, as already explained in the Introduction, was to equip the large millimetre telescope at Pico Veleta with an innovative deep-field camera. Even if the KID concept had been proposed more than five years earlier\cite{Day03}, at that time multiplexed KID hadn't been demostrated yet at a telescope, and the achieved sensitivities on the sky were still rather poor. Despite that, new ideas like the Lumped Element KID (LEKID) proposed by Doyle \textit{et al.}\cite{Doyle08} and the encouraging preliminary results concerning antenna-coupled KID \cite{Yates08} convinced us that building such an instrument was possible in a relatively short amount of time. Three European laboratories (Institut Neel, AIG Cardiff, SRON) decided on this base to start an intense phase of development culminated in October 2009, i.e. less than one year after the kick-off, in the first astronomical light for a multiplexed (MUX) KID camera. In the first observational run we demonstrated the performances of both antenna-coupled detectors and LEKID at 150 GHz\cite{Monfardini10}. From the MUX electronics point-of-view, we used during the observations both an FFTS\cite{Yates09} (Fast Fourier Transform Spectrometer) system and a more classical DDC\cite{Swenson09} (Direct Down Conversion). Despite the still poor sensitivities, the run was a big success from the technological point-of-view.

\begin{figure}
\begin{center}
\includegraphics[%
  width=0.50\linewidth,
  keepaspectratio]{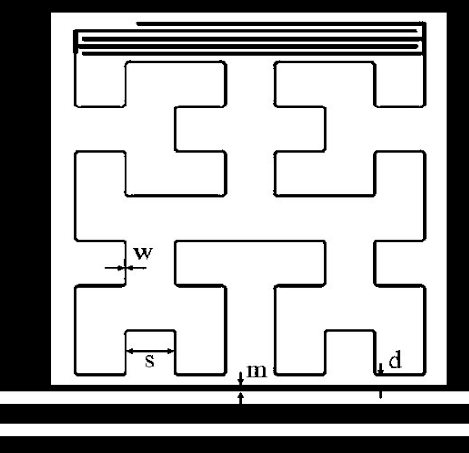}
\end{center}
\caption{The NIKA LEKID design based on a fractal-shaped inductor and an interdigited capacitor. The LEKID is inductively coupled to a CoPlanar Waveguide (CPW). Black: thin (e.g. 20 nm) Aluminium; white: Silicon substrate. The pixel size is about 2 mm. The main design parameters are shown in the drawing and explained in more datail in Roesch \textit{et al.}\cite{Roesch12}}
\label{Hilbert}
\end{figure}

In Run 2, in October 2010, NIKA wa improved to a dual-band\cite{Monfardini11} (150 GHz and 240 GHz) instrument. Also, it had grown already to a hundreds pixels fully multiplexed instrument. For that test, we adopted single polarisation LEKID at 150 GHz and antenna-coupled KID at 240 GHz. The sensitivity at 150 GHz improved by a factor of three compared to the previous experience. 

In Run 3, November 2011, we demonstrated for the first time a new LEKID design allowing to absorb both polarisations (see Fig.~\ref{Hilbert}). This design allowed to further improve the detectivity by a factor of two, and thus to reach sensitivities comparable, at that time, with the best instruments employing the more classical TES bolometers technology. From Run 3 on, NIKA has adopted LEKID for both frequencies. Another important achievement of Run 3 was the implementation of an innovative modulated read-out\cite{Calvo13} permitting to substantially improve the relative photometry performances of KID-based instruments. The flux stability measured on primary calibrators, i.e. planets, is better than 10\%, comparable to existing instruments and more than three times better than during Run 2. The MUX electronics used for both Runs 2 and 3 was an open-source readout built around a ROACH board\cite{roach}. The hardware has been developed by a collaboration led by the University of California Santa Barbara (UCSB).

\begin{figure}
\begin{center}
\includegraphics[%
  width=0.75\linewidth,
  keepaspectratio]{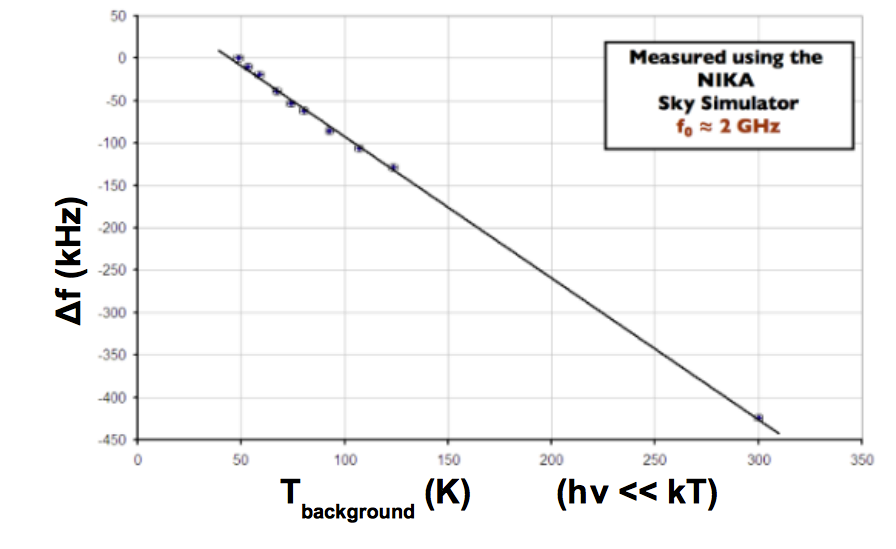}
\end{center}
\caption{(color online) KID linearity demonstrated in laboratory under realistic conditions. Y-axis: frequency shift of the resonance (KID measured signal), or pixel (kHz); x-axis: optical background temperature (K). The solid line represents the linear fit of the experimental points.}
\label{linearity}
\end{figure}

During the technical Run 4, in June 2012, NIKA was permanently installed in the telescope receivers cabin with a completely new re-imaging optics. 

In Run 5, November 2012, we used for the first time the intrinsic KID linearity (see Fig.~\ref{linearity}) over large power range to implement a real-time opacity calculator\cite{Catalano13}. 

The last (so far) Run 6, in June 2013, was mainly dedicated to demonstrate improved sensitivity (by roughly 30\% compared to runs 3-4-5) and NIKA successful integration in the IRAM observatory software envinronment. 

Integrating the six Runs, NIKA has been on the sky at the 30-m telescope for about four weeks only. That relatively short time has been enough to develop, starting from a previously non-demonstrated technology, a funtional instrument with state-of-the-art performances. In the next paragraph we present a few selected results achieved in the most recent runs 5 and 6.

\section{Latest NIKA results and plans for 2013-2015}

\begin{figure}
\begin{center}
\includegraphics[%
  width=0.75\linewidth,
  keepaspectratio]{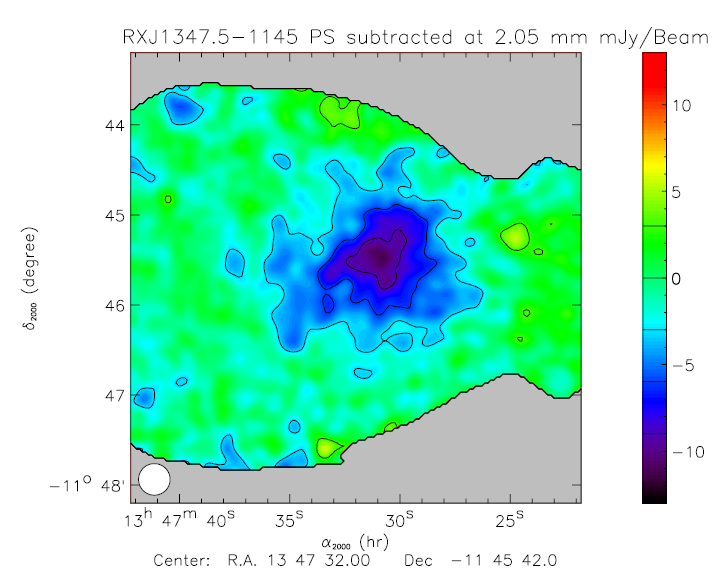}
\end{center}
\caption{(color online) First detection of the SZ effect using a KID camera. On-the-fly (zig-zag) map taken during the NIKA run 5 in November 2012. Target: cluster RXJ1347.5-1145.}
\label{SZ}
\end{figure}

During the scientific run 5, NIKA detected, for the first time using KID, two clusters of galaxies via the Sunayev-Zeldovich (SZ) effect. In SZ, the Cosmic Microwave Background (CMB) photons interact through the inverse Compton mechanism with the hot electrons in the intergalactic plasma. This interaction determines a spectral deformation of the black-body (2.7 K) CMB radiation. In particular, around 150 GHz this results in a small (e.g. 0.1\% or lower) dip of the CMB flux. We present a preliminary map in Fig.~\ref{SZ}; for more detailed results please refer to the paper in preparation by R. Adam \textit{et al.}\cite{Adam13}.

\begin{figure}
\begin{center}
\includegraphics[%
  width=0.95\linewidth,
  keepaspectratio]{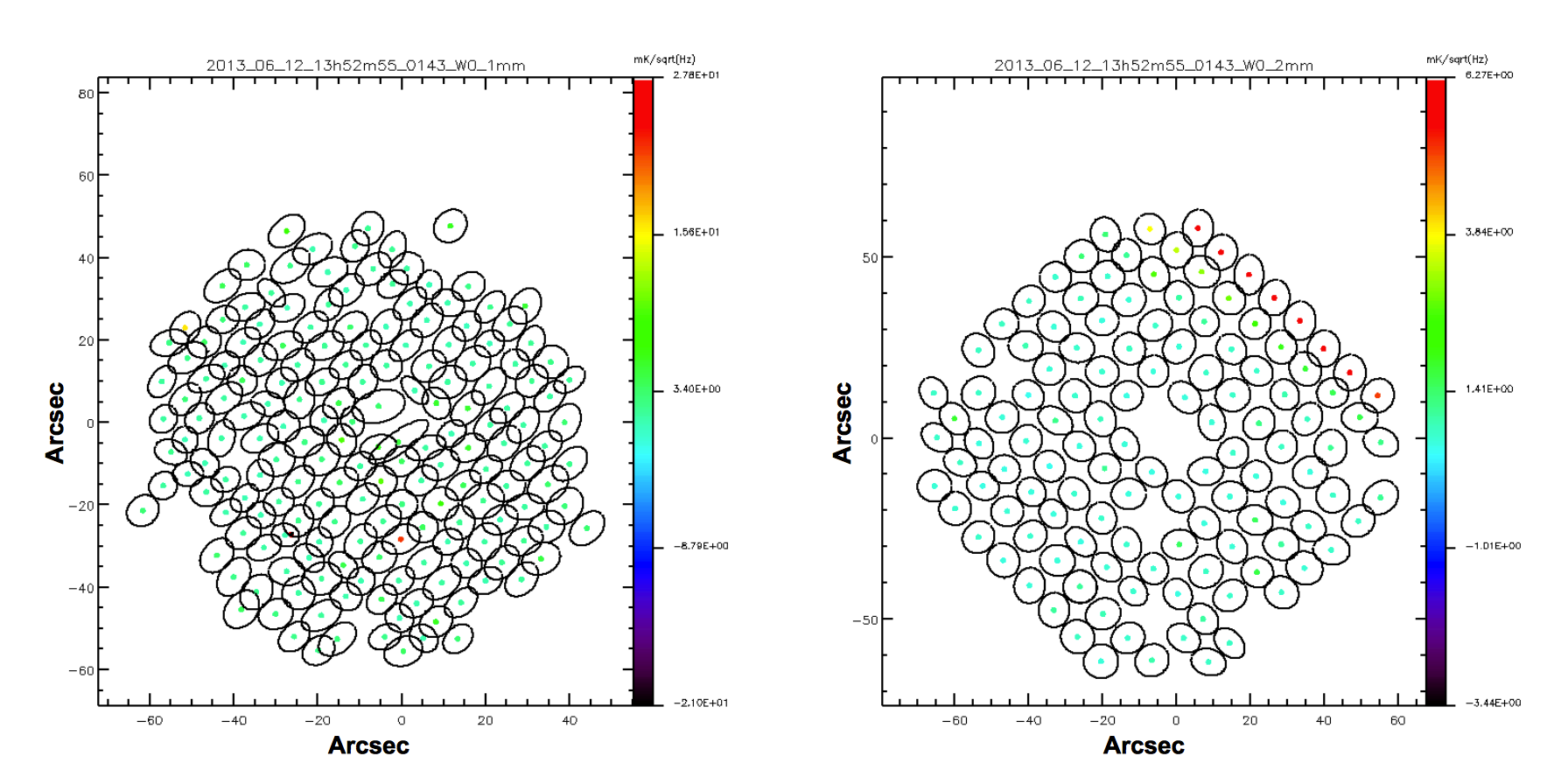}
\end{center}
\caption{(color online) Projected beams measured during run6. Left: 240 GHz (1.25 mm) band; about 190 valid pixels (design: 224). Right: 150 GHz (2 mm); 124 valid pixels (design: 132).}
\label{Geometry}
\end{figure}

In the Commissioning run 6, in June 2013, we have substantially improved the cosmetics (e.g. dead pixels, uniformity) of our arrays, and in particular the 150 GHz one, by suppressing unwanted RF propagation modes in the CPW readout line. This also allowed to optimise the resonators coupling factor to the same readout line (Q$_c \approx 1\div2\cdot10^4$ for the NIKA arrays) in order to achieve the best signal-to-noise. The asymmetric modes are reduced by strapping the ground plane across the readout line. Moreover, we have optimised the Aluminium film thickness and deposition techniques. The distribution of the beams, projected on the Sky, is shown in Fig.~\ref{Geometry}. The 150 GHz array has, among the 132 design pixels, about 94\% good beams. On the 240 GHz focal plane, 85\% of the beams have been identified.

\begin{figure}
\begin{center}
\includegraphics[%
  width=1.05\linewidth,
  keepaspectratio]{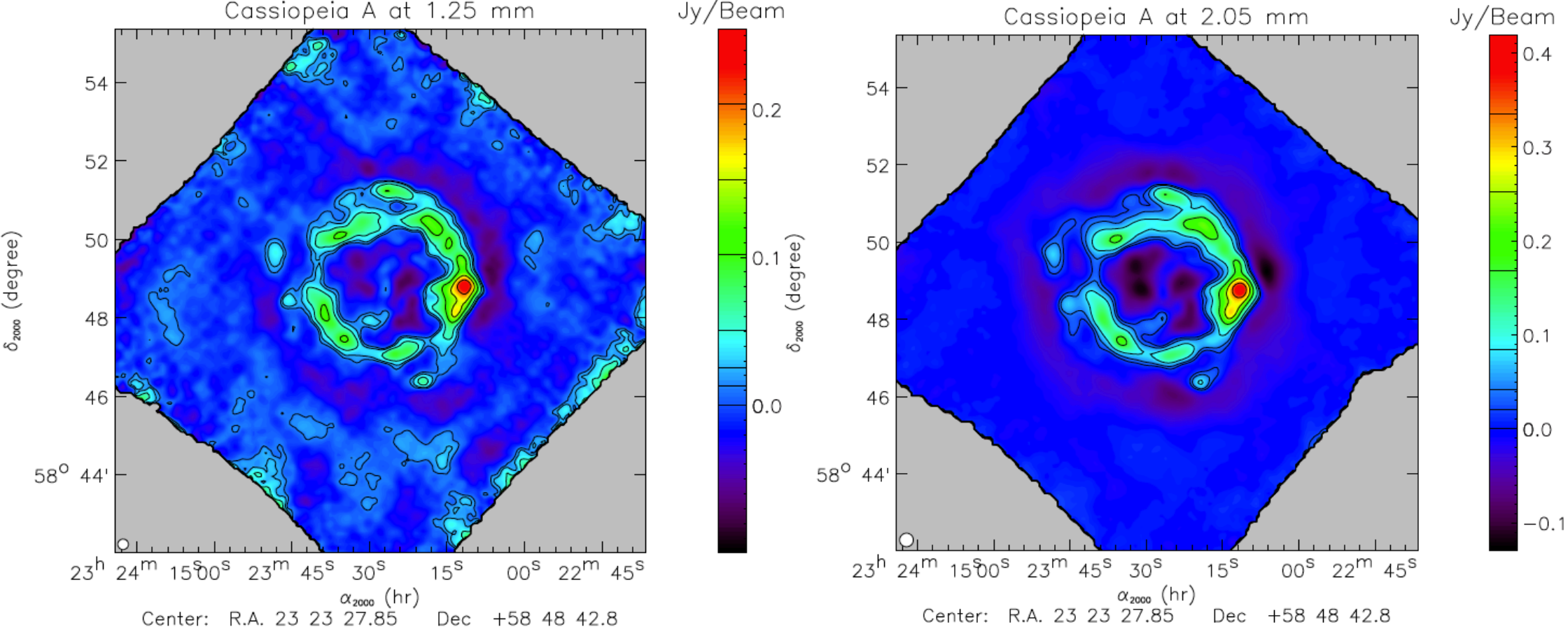}
\end{center}
\caption{(color online) Cassiopea A SNR imaged by NIKA during run 6 in June 2013. This map is to be considered a quick-look result. These pictures are shown to demonstrate the functionality of the NIKA camera for extended sources imaging. Left: 240 GHz (1.25 mm) band; right: 150 GHz (2 mm).}
\label{CasA}
\end{figure}

In the Fig.~\ref{CasA} we present an example map taken during run 6. The quick-look, filtered image refers to the young (330 years) supernova remnant (SNR) Cassiopea A. More sophisticated analysis on this and other extended sources is ongoing and the results will be reported in more specific publications.

The average Noise Equivalent Flux Density (NEFD), or weak-source flux detectivity, achieved by NIKA are $40 \mathrm{mJy}\cdot s^{1/2}$ and $15 \mathrm{mJy}\cdot s^{1/2}$ per beam respectively at 240 GHz and 150 GHz. These sensitivities are calculated for a telescope elevation of 60 deg and an atmosphere opacity equivalent to $\tau_{225GHz} = 0.1$. The best pixels at 150 GHz show a sensitivity below $10\mathrm{mJy}\cdot s^{1/2}$ per beam, indicating that the KID perform already in line with the state-of-the-art bolometers\cite{gismo13}. In terms of NEP, the NIKA 150 GHz pixels lie in the high-$10^{-17} W\cdot Hz^{-1/2}$ range, as expected for the best ground-based observations in this band of frequencies.

\section{The next generation: NIKA-2 (2015-2025)}

NIKA, with its $\approx$ 350 pixels, was intented since the beginning to pave the way for the next generation KID camera at the 30-m telescope. 

\begin{figure}
\begin{center}
\includegraphics[%
  width=1.05\linewidth,
  keepaspectratio]{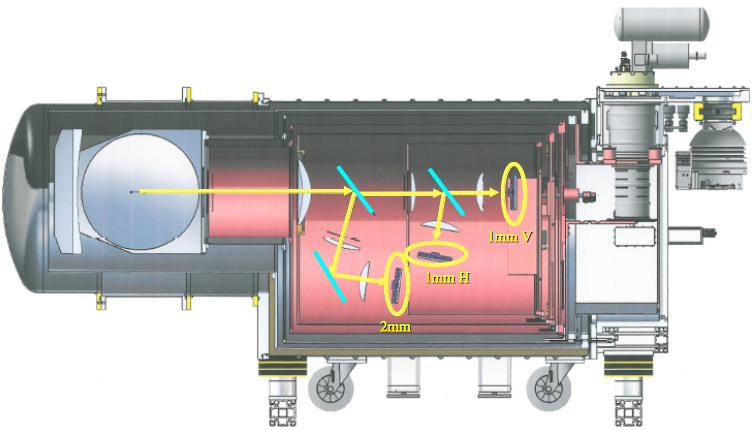}
\end{center}
\caption{(color online) NIKA-2 instrument. The beam splitting is achieved by a dichroic first and a polariser (yellow line). The three KID arrays are cooled down at roughly 100 mK by means of an ad-hoc dilution fridge. Total length: 2.3 m; weight: 980 kg.}
\label{NIKA2}
\end{figure}

NIKA-2 will fully sample the telescope extended field-of-view of 6.5 arc-min using about 5,000 pixels spread over three KID arrays. Besides dual-band imaging capabilities (still 150 GHz and 240 GHz), it will be able to measure the linear polarization of the targeted sources at 240 GHz. That is achieved with double beam splitting (dichroic + polarizer) at low temperature (see Fig.~\ref{NIKA2}) and a rotating halfwave plate at 300 K in front of the cryostat window.  

The large NIKA-2 cryostat is, as already the case for NIKA, cryogen-free. The 4 K cold point is assured by two pulse-tubes providing together 2 W cooling power. An ad-hoc dilution insert developed at the Institut N\'eel allows cooling down the detectors and part of the optics at 100 mK.

The first functional prototype of the NIKA-2 150 GHz array has been fabricated and pre-tested. It comprises four readout lines and 1,020 pixels, for a fully filled focal plane diameter of 80 mm. The detectors are realised from a single Aluminium (t=$12\div25$ nm) film on a 100 mm Silicon wafer. An AR (Anti-Reflecting) treatment is applied by micromachining or dicing to the back of the wafer to improve the spectral response flatness of the LEKID. The exact design and the fabrication details are entirely derived from the NIKA developments.

\begin{figure}
\begin{center}
\includegraphics[%
  width=0.75\linewidth,
  keepaspectratio]{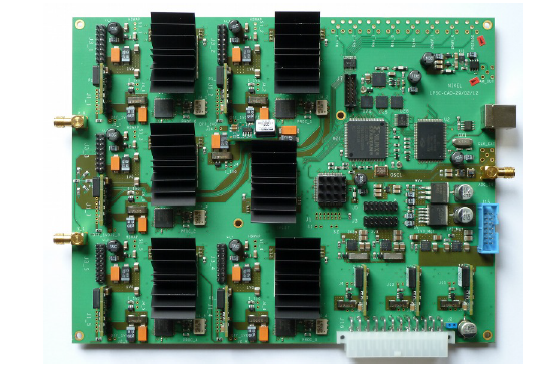}
\end{center}
\caption{(color online) NIKEL board equipped with 12-bit 1 GSPS ADC and 16-bit 1 GSPS DAC. NIKEL allows to multiplex up to 400 pixels over 500 MHz bandwidth with the large ADC dynamics needed by ground-based applications. }
\label{NIKEL}
\end{figure}

The MUX electronics for NIKA-2 is the NIKEL system described in more detail by O. Bourrion \textit{et al.}\cite{Bourrion11}. NIKEL has been specifically realised for NIKA and NIKA-2 and allows a multiplexed readout of up to 400 channels/pixels over 500 MHz bandwidth. It generates the modulated signal needed by the RF synthesizers, averages the I,Q data down to the final rate ($\approx$25 Hz) and diffuses them by UDP (User Datagram Protocol ) packets over a local network to the acquisition computers.

\section{Conclusions}
The NIKA dual-band continuum instrument is now fully operational and will be open to the 30-m telescope observers from the next Winter (2013/14). In six technical and scientific (restricted to the NIKA collaboration) runs we demonstrated competitive sensitivities at both 150 GHz and 240 GHz and good photometry performances. Moreover, thanks to the intrinsic KID linearity we provide the astronomers with a real-time estimation of the atmospheric opacity correction. This correction is calculated along the line-of-sight and  for the real detectors band. It is thus in the long term more reliable than the IRAM tipping radiometer which works at 225 GHz at a fixed azimuth direction.

The next generation NIKA-2 camera will cover a larger field-of-view (6.5 arc-min compared to 2 arc-min in NIKA), preserve dual-band imaging capabilities and measure in addition the linear polarisation at 240 GHz. This is achieved with three large LEKID arrays (sensitive area diameter = 80 mm) and a total pixels count of 5,000. NIKA-2, being fabricated in Grenoble, is officially selected by IRAM as the next generation continuum instrument at the 30-m telescope. It will be installed for Commissioning in 2015. 

\begin{acknowledgements}
This work has been partially funded by the Foundation Nanoscience Grenoble, the ANR 
under the contracts "MKIDS" and "NIKA". This work has been partially supported by the LabEx FOCUS ANR-11-LABX-0013. This work has benefited from the support of the European Research Council Advanced Grant ORISTARS under the European Union's Seventh Framework Programme (Grant Agreement no. 291294). We acknowledge the crucial contributions of ex-members of the NIKA collaboration, in particular Loren Swenson, Markus Roesch, Angelo Cruciani, Julien Minet. We thank all the IRAM Granada staff for the excellent support before, during and after NIKA observations. The NIKA cryostat and the readout electronics have been fabricated by the Cryogenics and Electronics groups in Grenoble.
\end{acknowledgements}

\pagebreak

\end{document}